%% file: main.tex
\documentclass[12pt,english]{article}

\usepackage{authblk}
\usepackage{amsmath,amsfonts,mathtools}
\usepackage{bbm,bm}
\usepackage{setspace}
\usepackage{lineno}

\usepackage{graphicx}
\usepackage{float}

\usepackage[colorinlistoftodos]{todonotes}

\usepackage{xcolor,colortbl}
\usepackage{framed}
\colorlet{shadecolor}{lightgray!20}

\usepackage{fullpage}
\usepackage[
  a4paper,
  left=2.5cm,
  right=2.5cm,
  top=2.5cm,
  bottom=2.5cm
]{geometry}

\usepackage{tikz}
\usetikzlibrary{
  fit,
  calc,
  decorations.pathreplacing,
  arrows.meta,
  backgrounds
}
\tikzset{>={Latex[width=1.5mm,length=1.5mm]}}

\usepackage{media9}
\usepackage[labelfont=bf]{caption}

\RequirePackage[
  backend=biber,
  style=authoryear,
  natbib=true,
  maxcitenames=2,
  uniquename=false,
  hyperref=true,
  url=true
]{biblatex}

\usepackage[
  hypertexnames=false
]{hyperref}

\newenvironment{content}{
  \setlength{\parskip}{10pt}
}{}

\title{
Non-Homogeneous Markov-Switching Generalized Additive Models for Location, Scale, and Shape
}

\author{Katharina Ammann$^{1}\footnote{Corresponding author; email: \texttt{katharina.ammann@uni-bielefeld.de}.}$, Timo Adam$^1$, Jan-Ole Koslik$^1$\\ $^1$Bielefeld University, Germany}

\date{\vspace{-5ex}} 

\begin{document}

\pagenumbering{arabic}
\setcounter{page}{1}

\maketitle
\input{content/00_abstract}

\begin{content}

\input{content/01_introduction}
\input{content/02_methods}
\input{content/03_simulation}
\input{content/04_application}
\input{content/05_discussion}

\appendix
\renewcommand{\thesection}{Appendix~\Alph{section}}

\input{content/XX_Appendix}

\end{content}

\printbibliography

\end{document}

%% file: content/00_abstract.tex
\begin{abstract}
\text{We} propose an extension of Markov-switching generalized additive models for location, scale, and shape (MS-GAMLSS) that allows covariates to influence not only the parameters of the state-dependent distributions but also the state transition probabilities. Traditional MS-GAMLSS, which combine distributional regression with hidden Markov models, typically assume time-homogeneous (i.e., constant) transition probabilities, thereby preventing regime shifts from responding to covariate-driven changes. Our approach overcomes this limitation by modeling the transition probabilities as smooth functions of covariates, enabling a flexible, data-driven characterization of covariate-dependent regime dynamics. Estimation is carried out within a penalized likelihood framework, where automatic smoothness selection controls model complexity and guards against overfitting. We evaluate the proposed methodology through simulations and applications to daily Lufthansa stock prices and Spanish energy prices. Our results show that incorporating macroeconomic indicators into the transition probabilities yields additional insights into market dynamics.
Data and \texttt{R} code to reproduce the results are available online.
\end{abstract}

\noindent \textbf{Keywords:} distributional regression models; generalized additive models for location, scale, and shape; hidden markov models; time series modeling.

%% file: content/01_introduction.tex
\section{Introduction}
\vspace{-\baselineskip}
In recent years, Markov-switching generalized additive models for location, scale, and shape (MS-GAMLSS) have emerged as a flexible class of time series regression models that capture dynamic structural changes in the conditional response distribution through a latent state process \parencite{Adam2022}.
Traditionally, the transition probabilities that govern that state process are assumed to be time-homogeneous, i.e., constant over time. Although this assumption simplifies estimation, it may not adequately reflect real-world applications. For example, in financial markets, transitions between calm and volatile regimes are often driven by macroeconomic conditions rather than occurring purely at random \parencite{foroni2024expectile}. In this paper, we propose an extension of time-homogeneous MS-GAMLSS in which the transition probabilities between latent states depend on covariates. This suggested model formulation makes it possible to represent scenarios in which the dynamics of the latent states are systematically influenced by observable conditions.

The proposed methodology builds on a rich literature.
Since their introduction by \citet{Rigby2005}, GAMLSS have transformed regression modeling by allowing not only the mean but also other parameters of the response distribution --- such as variance, skewness, and kurtosis --- to depend on covariates via smooth functions. Applications span a wide range of fields, including flood risk modeling \parencite{Villarini2009}, centile estimation in clinical data \parencite{Hossain2016}, speech intelligibility testing \parencite{Hu2015, Stasinopoulos2024}, and prediction of birthweight from ultrasound measures \parencite{Stasinopoulos2024}. Further use cases include demand forecasting in e-commerce \parencite{Ulrich2021}, exposure-at-default modeling \parencite{Wattanawongwan2023}, asset pricing under distributional uncertainty \parencite{Regis2023}, and modeling election results \parencite{balzer2025gradient}.

The flexibility of GAMLSS has also attracted the interest of major institutions. For example, the International Monetary Fund incorporated GAMLSS into its stress-testing framework for the U.S.\ economy, yielding insights into systemic financial risks and informing policy recommendations in collaboration with the U.S.\ Treasury, the Federal Reserve, and other key agencies \parencite{IMF2015}. Similarly, the Bank of England applied GAMLSS to assess vulnerabilities in the UK mortgage market, with a particular focus on tail risk and distributional shifts in borrower characteristics \parencite{BankEngland2017}. In the private sector, institutions such as the Bank of America used GAMLSS to benchmark internal credit risk models, especially for components like loss-given-default and exposure-at-default, where flexible modeling of skewed and heavy-tailed loss distributions is essential \parencite{BoA2020}.

While most applications of GAMLSS assume independent observations, this assumption is often violated in time series data, where temporal dependence and regime-switching behavior are common \parencite{GoldfeldQuandt1973, Hamilton1989}. To address this, \citet{Langrock2018} introduced MS-GAMLSS, a class of time series regression models that combines GAMLSS with hidden Markovian state structures. MS-GAMLSS extend the GAMLSS framework to time series by allowing the parameters of the response distribution to vary with covariates via smooth functions depending on an underlying, unobserved state. In other words, the time series is assumed to be generated by separate distributional regression models, with each model corresponding to a different regime.
Related contributions include \citet{Hambuckers2018}, who incorporated compound Poisson regression into Markov-switching models, and \citet{Adam2022}, who proposed gradient boosting within MS-GAMLSS to accommodate high-dimensional predictor settings. Building on this line of research, the present work further extends MS-GAMLSS by explicitly modeling the hidden state process as a covariate-dependent, and thus time-non-homogeneous, Markov chain, where covariates directly influence the transition probabilities.

The remainder of the paper is structured as follows. Section~\ref{sec:methods} introduces the proposed methodology and explains how it models state-specific distributional dynamics and covariate-dependent state transitions. In Section~\ref{sec:simulation}, we evaluate the model’s performance in a simulation study. 
Two case studies, involving daily closing prices of Lufthansa stock and energy prices in Spain, are given in Section~\ref{sec:application}. Data and \texttt{R} code to reproduce the results are available on GitHub\footnote{\texttt{https://github.com/ammann99/nonhomogeneousMSGAMLSS}.}. 

%% file: content/02_methods.tex
\section{Methods}
\label{sec:methods}
\vspace{-\baselineskip}
\subsection{Model Formulation and Dependence Structure}
\label{subsec:model_formulation}
\begin{figure}
\centering
\begin{tikzpicture}[node distance = 2cm]
\tikzset{state/.style = {circle, draw, minimum size = 42pt, scale = 0.82}}

\node [state,fill=white!10] (4) [] {$S_{t-2}$};
\node [state,fill=white!10] (6) [right = 5mm of 4] {$S_{t-1}$};
\node [state,fill=white!10] (5) [right = 5mm of 6] {$S_{t}$}; 
\node [state,fill=white!10] (7) [right = 5mm of 5] {$S_{t+1}$};

\node [state,fill=white!10 ] (8) [above = 5mm of 6] {$Y_{t-1}$};
\node [state,fill=white!10 ] (9) [above = 5mm of 5] {$Y_{t}$}; 
\node [state,fill=white!10 ] (10) [above =5mm of 7] {$Y_{t+1}$};
\node [state,fill=white!10 ] (22) [above = 5mm of 4] {$Y_{t-2}$};

\node [state,fill=white!10] (15) [left = 5mm of 4] {$\cdots$};
\node [state,fill=white!10] (16) [right = 5mm of 7] {$\cdots$};

\node [state,fill=white!15] (z1) [below = 5mm of 4] {$\mathbf{z}_{t-2}$};
\node [state,fill=white!15] (z2) [right = 5mm of z1] {$\mathbf{z}_{t-1}$};
\node [state,fill=white!15] (z3) [right = 5mm of z2] {$\mathbf{z}_{t}$}; 
\node [state,fill=white!15] (z4) [right = 5mm of z3] {$\mathbf{z}_{t+1}$};

\node [state,fill=white!10] (19) [above = 5mm of 8] {$\mathbf{x}_{t-1}$};
\node [state,fill=white!10] (20) [above = 5mm of 9] {$\mathbf{x}_{t}$};
\node [state,fill=white!10] (21) [above = 5mm of 10] {$\mathbf{x}_{t+1}$};
\node [state,fill=white!10] (23) [above = 5mm of 22] {$\mathbf{x}_{t-2}$};

\draw[->, black, line width=0.2pt] (15) to (4);
\draw[->, black, line width=0.2pt] (7) to (16);
\draw[->, black, line width=0.2pt] (4) to (6);
\draw[->, black, line width=0.2pt] (6) to (5);
\draw[->, black, line width=0.2pt] (5) to (7);
\draw[->, black, line width=0.2pt] (4) to (22);
\draw[->, black, line width=0.2pt] (6) to (8);
\draw[->, black, line width=0.2pt] (5) to (9);
\draw[->, black, line width=0.2pt] (7) to (10);
\draw[->, black, line width=0.2pt] (23) to (22);
\draw[->, black, line width=0.2pt] (19) to (8);
\draw[->, black, line width=0.2pt] (20) to (9);
\draw[->, black, line width=0.2pt] (21) to (10);

\draw[->, black, line width=0.2pt] (z1) -- (4);
\draw[->, black, line width=0.2pt] (z2) -- (6);
\draw[->, black, line width=0.2pt] (z3) -- (5);
\draw[->, black, line width=0.2pt] (z4) -- (7);

\draw[decorate,decoration={brace,amplitude=10pt}] 
  (7.5,4.0) -- (7.5,2.7) 
  node[midway,xshift=10pt,yshift=0pt,right] 
  {\colorbox{white!10}{\strut\shortstack[l]{covariates}}};

\draw[decorate,decoration={brace,amplitude=10pt}] 
  (7.5,2.3) -- (7.5,1.0) 
  node[midway,xshift=10pt,yshift=0pt,right] 
  {\colorbox{white!10}{\strut response variable}};

\draw[decorate,decoration={brace,amplitude=10pt}] 
  (7.5,0.6) -- (7.5,-0.7) 
  node[midway,xshift=12pt,yshift=1pt,right] {hidden states};

\draw[decorate,decoration={brace,mirror,amplitude=10pt}] 
  (7.5,-2.4) -- (7.5,-1.1) 
  node[midway,xshift=10pt,yshift=0pt,right] 
  {\colorbox{white!15}{\strut covariates}};

\end{tikzpicture}

\caption[Illustration of the dependence structure of a non-homogeneus MS-GAMLSS.]{\textbf{Illustration of the dependence structure of a non-homogeneus MS-GAMLSS.} Covariates influence both the hidden state process \(\{S_t\}_{t=1,\ldots,T}\) and the observed state-dependent process \(\{Y_t\}_{t=1,\ldots,T}\).}
\label{fig:dep_struc}
\end{figure}
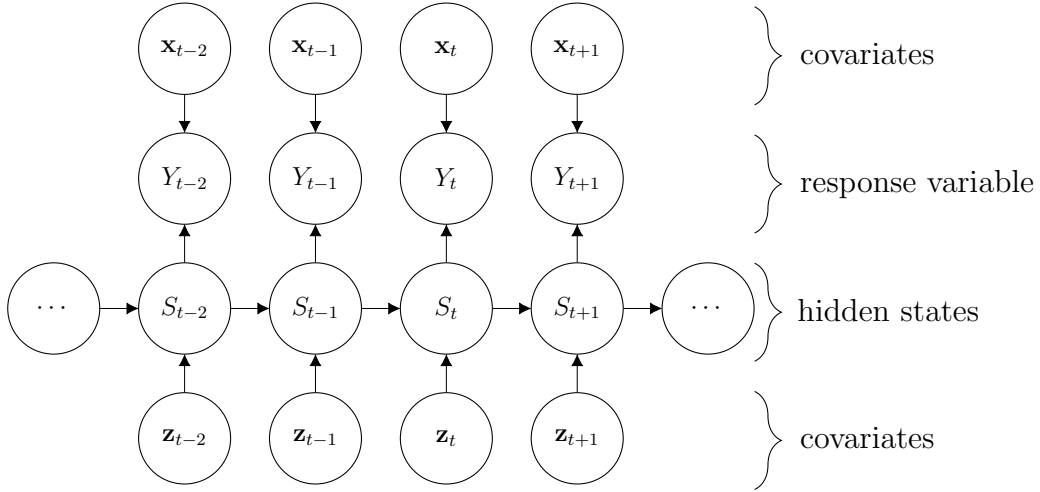

MS-GAMLSS combine a hidden Markov chain, representing latent regimes, with an observed state-dependent response process (see Fig. \ref{fig:dep_struc} for an illustration of the dependence structure). In this subsection, we introduce the structure and covariate-dependent transitions of the proposed non-homogeneous MS-GAMLSS.

The \textit{hidden state process} is represented by a discrete-time, \( N \)-state Markov~chain \(\{S_t\}_{t=1,\ldots,T}\). A first-order Markov property is assumed, i.e., 
\[
\Pr(S_{t+1}\mid S_1,\dots,S_t) = \Pr(S_{t+1}\mid S_t), \quad t=1,\dots,T-1.
\] 
This simplifies the dependence structure and allows efficient likelihood computation, while higher-order extensions are possible if necessary \parencite{zucchini2016hidden}.

The initial state probabilities, denoting the probability of starting in each state at time \( t = 1 \), are summarized in the vector \(\boldsymbol{\delta}^{(1)} = (\delta_1,\ldots,\delta_N)\), with \(\delta_i = \Pr(S_1 = i)\). For time-homogeneous Markov chains, state transitions are characterized by a time-invariant \( N \times N \) transition probability matrix (t.p.m) \(\boldsymbol{\Gamma} = (\gamma_{i,j})_{i,j=1, \dots, N} \), where $\gamma_{i,j} = \Pr(S_{t+1}=j \mid S_t=i), \text{ for } i, j = 1, \ldots, N$. 

To allow covariates to affect regime-switching dynamics, we consider non-homogeneous (i.e., time-varying) transition probabilities. These are described by the time-dependent t.p.m. \(\boldsymbol{\Gamma}_t =(\gamma^{(t)}_{i,j})_{i,j=1,\dots,N}\), with
\[
\gamma^{(t)}_{i,j} = \Pr(S_{t+1}=j \mid S_t=i, \mathbf{z}_t), \text{ for } i,j=1,\dots,N,
\]
where $\mathbf{z}_t = (z_{1t}, \dots, z_{Kt})^\top$ represents covariates influencing state transitions.

A multinomial logistic link function is used to link the additive predictor to the transition probabilities, i.e.,
\[
\gamma^{(t)}_{i,j} = \frac{\exp\bigl(\eta_{i,j}^{(t)}(\mathbf{z}_t)\bigr)}{\sum_{l=1}^N \exp\bigl(\eta_{i,l}^{(t)}(\mathbf{z}_t)\bigr) }, \quad \text{for }i,j = 1, \dots, N,
\]
where for $i \neq j$ the linear predictors take the form
{
\setlength{\abovedisplayskip}{6pt}
\setlength{\belowdisplayskip}{6pt}
\[
\eta_{i,j}^{(t)}(\mathbf{z}_t) = \beta_0^{(i,j)} + \sum_{k=1}^K \beta_k^{(i,j)} z_{kt} + \sum_{k=1}^K s_k^{(i,j)}(z_{kt}),
\]
}

\noindent and $\eta_{i,i}^{(t)}(\mathbf{z}_t) = 0$ for identifiability. $\beta_0^{(i,j)}$ is a state-pair-specific intercept, $\beta_k^{(i,j)}$ are linear coefficients, and $s_k^{(i,j)}(\cdot)$ are smooth functions modeled using spline basis functions (see Equation (\ref{eqn:smooth_representation})).

The observed \textit{state-dependent process} $\{Y_t\}_{t=1, \dots, T}$ is assumed to be conditionally independent, given the states, and to follow some parametric distribution $\mathcal{D}$,
\begin{align}
Y_t \sim \mathcal{D}\bigl(\mu_t^{(s_t)}, \sigma_t^{(s_t)}, \nu_t^{(s_t)}, \tau_t^{(s_t)}\bigr),
\label{distribution}
\end{align}
where $\mu_t^{(s_t)}$, $\sigma_t^{(s_t)}$, $\nu_t^{(s_t)}$, and $\tau_t^{(s_t)}$ are state-specific, covariate-dependent parameters representing the mean, variance, skewness, and kurtosis, respectively. These parameters are linked to additive predictors via monotonic link functions \(g_k(\cdot)\), such that
{
\allowdisplaybreaks
\[
\begin{aligned}
g_1\bigl(\mu_t^{(i)}\bigr) &= \mathbf{X}_1 \boldsymbol{\beta}_1^{(i)} + \sum_{j=1}^{J_1} s_{1,j}^{(i)}(x_{j,t}), \\
g_2\bigl(\sigma_t^{(i)}\bigr) & = \mathbf{X}_2 \boldsymbol{\beta}_2^{(i)} + \sum_{j=1}^{J_2} s_{2,j}^{(i)}(x_{j,t}), \\
g_3\bigl(\nu_t^{(i)}\bigr) &= \mathbf{X}_3 \boldsymbol{\beta}_3^{(i)} + \sum_{j=1}^{J_3} s_{3,j}^{(i)}(x_{j,t}), \\
g_4\bigl(\tau_t^{(i)}\bigr) &= \mathbf{X}_4 \boldsymbol{\beta}_4^{(i)} + \sum_{j=1}^{J_4} s_{4,j}^{(i)}(x_{j,t}),
\end{aligned}
\]
}
where \(\mathbf{X}_k\) denote design matrices for the linear effects, \(\boldsymbol{\beta}_k^{(i)}\) state-specific linear regression coefficients, and \(s_{k,j}^{(i)}(\cdot)\) smooth covariate effects. The number of covariates or smooth terms in each additive predictor is denoted by \(J_k\). Each distribution parameter is transformed using a monotonic link function $g_k(\cdot)$ to map it onto the real-valued additive predictors, ensuring parameter constraints (e.g., positivity for the variance).

All smooth effects in the state and state-dependent process are represented as finite linear combinations of $M$ fixed spline basis functions, i.e.,
\begin{align}
\label{eqn:smooth_representation}
s(x) = \sum_{m=1}^M b_m \phi_m(x),
\end{align}
with spline coefficients \(b_{m}\) and spline basis functions $\phi_m(x)$ (\citealp{de1978practical, eilers1996flexible}).

This flexible model structure enables state-dependent and covariate-driven distributional regression for time series data.
Although the conditional independence assumption of the observations, given the states might seem restrictive, it is supported by the fact that serial correlation is explicitly that the state process induces serial dependence.
\vspace{-\baselineskip}

\subsection{Likelihood Evaluation and Model Fitting}
\label{sec:2.2}

Since the likelihood involves summing over all possible state sequences, a direct evaluation quickly becomes infeasible. Instead, likelihood evaluation in Markov-switching models can conveniently be carried out using the forward algorithm, which employs the so-called \textit{forward variables} to recursively compute the likelihood (\citealp{zucchini2016hidden}). The forward variables $\alpha_t(j)$ represent the joint density of the observations up to time $t$ and being in state $j$ at time $t$, defined as 
 $  \alpha_t(j) = f(y_1, \dots, y_t, S_t = j)$. 
The forward variables can be grouped into a vector 
$
\boldsymbol{\alpha}_t = (\alpha_t(1), \alpha_t(2), \dots, \alpha_t(N)).
$

The key idea of the forward algorithm is that these forward variables can be updated recursively, thus enabling likelihood computation in linear time.
At $t = 1$, the forward variables are initialized as 
{
\setlength{\abovedisplayskip}{6pt}
\setlength{\belowdisplayskip}{6pt}
\begin{align*}
\boldsymbol{\alpha}_1 = \boldsymbol{\delta}^{(1)} \mathbf{P}(y_1),
\end{align*}
}

\noindent where $ 
\mathbf{P}(y_t) = \text{diag}(
f(y_t \, \vert \, \mu_t^{(1)}, \sigma_t^{(1)}, \nu_t^{(1)}, \tau_t^{(1)}),
\, \ldots \,,
f(y_t \, \vert \, \mu_t^{(N)}, \sigma_t^{(N)}, \nu_t^{(N)}, \tau_t^{(N)})
)
$ is a diagonal matrix containing the state-dependent response densities. 

For subsequent time steps, $t = 2, \dots, T$, the forward variables are updated iteratively using the recursion 
{
\setlength{\abovedisplayskip}{6pt}
\setlength{\belowdisplayskip}{6pt}
\begin{align*}
\boldsymbol{\alpha}_t = \boldsymbol{\alpha}_{t-1} \boldsymbol{\Gamma}^{(t)} \mathbf{P}(y_t).
\end{align*}}
The likelihood then follows from the law of total probability as the sum of the forward probabilities at time $T$,
\vspace{-0.2cm}
{
\setlength{\abovedisplayskip}{-3pt}
\setlength{\belowdisplayskip}{3pt}
\begin{align*}
\mathcal{L}(\boldsymbol{\theta}) &= f(y_1, \dots, y_T) = \sum_{i=1}^N \alpha_T(i) = \boldsymbol{\alpha}_T \boldsymbol{1}^\top,
\end{align*}}
where $\boldsymbol{\alpha}_T$ is the vector of forward variables at time $T$, and $\boldsymbol{1}^\top$ is a row vector of ones.

To address numerical issues during the computation of the likelihood for long time series, such as underflow or overflow, we compute the log-likelihood using a scaling strategy, which employs \textit{standardized forward variables} normalized to avoid numerical instability (\citealp{lystig2002}; see Appendix~\ref{sec:app_stand_forw_var} for details about the scaled forward algorithm).

To prevent overfitting, we constrain the flexibility of the smooth functions in Equation \eqref{eqn:smooth_representation} by penalizing the associated coefficients. 
Let $\bm{b}_p$ denote the spline coefficient vector for a single smooth function. To ease notation, we use a single index. Assume $P$ smooth functions in total for the state-dependent distribution parameters and for the t.p.m.

Then, the model can be fitted by maximizing
{
\setlength{\abovedisplayskip}{3pt}
\setlength{\belowdisplayskip}{3pt}
\begin{align}
\label{eqn:penalized_llk}
\log \mathcal{L}(\boldsymbol{\theta}) - \frac{1}{2} \sum_{p = 1}^P \lambda_p \bm{b}_p^\top \bm{S}_p \bm{b}_p,
\end{align}}

\noindent where $\log \mathcal{L}(\boldsymbol{\theta})$ is evaluated using the scaled forward algorithm, $\lambda_1, \dotsc, \lambda_P$ are smoothing parameters and $\bm{S}_1, \dotsc, \bm{S}_P$ are penalty matrices for each smooth function.
For a simple \textit{P-spline} penalty \parencite{eilers1996flexible}, each $\bm{S}_p$ could be chosen as $\bm{D}^\top \bm{D}$, with $\bm{D}$ being the second-order difference matrix. A more general method for constructing these matrices is explained in detail by \citet{wood2017gam}. 
Clearly, each $\lambda_p$ determines the smoothness of the $p$-th smooth function. It is unconstrained for $\lambda_p = 0$, while converging to a linear function for $\lambda_p \to \infty$.
In practice, both the basis function evaluations in Equation \eqref{eqn:smooth_representation} and the penalty matrices $\bm{S}_p$ can conveniently be obtained from the \texttt{mgcv} \texttt{R} package \parencite{wood2017gam} based on model formulas, thereby offering great flexibility regarding the particular basis choice.

Automatic, data-driven smoothness selection is here conducted using approximate \textit{restricted maximum likelihood}, as introduced for Markov-switching models by \citet{koslik2024efficient}. The core idea of this approach is that the penalty in Equation \eqref{eqn:penalized_llk} imposes an improper multivariate Gaussian distribution for each $\bm{b}_p$ which motivates treating it as a random effect. Specifically, Equation \eqref{eqn:penalized_llk} implies that $\bm{b}_p \sim \mathcal{N}(0, \bm{S}_p^- / \lambda_j)$, where $ \bm{S}_p^-$ is the Moore-Penrose inverse of $\bm{S}_p$.
Hence, $\bm{\lambda} = (\lambda_1, \dotsc, \lambda_P)$ can be estimated based on the restricted likelihood function, with $\bm{b}_1, \dotsc, \bm{b}_P$ and all fixed effects integrated out. 
Intuitively, this restricted likelihood measures the average likelihood achieved based on many draws of $\bm{b}_1, \dotsc, \bm{b}_P$ from their distributions. This restricted likelihood will only be large for an adequate degree of smoothness. If the $\lambda_p$ are very small the drawn smooth functions are very wiggly, producing some very high likelihood scores but also many low scores. If the $\lambda_p$ are very large, most drawn functions are very straight, hence most likelihood scores will be fairly small.

On the computational side of things, the intractable high-dimensional integral is approximated using the so-called \textit{Laplace approximation}, which is based on a second-order Taylor approximation of \eqref{eqn:penalized_llk} around its mode for a given $\bm{\lambda}$. Ultimately this reduces the problem to applying iterative updates to the smoothness parameters based on subsequent penalized fits.
This typically leads to a moderate number of penalized model fits (10-50) to find a suitable smoothing parameter vector. Compared to naive alternatives using grid search, this is a much lower cost, especially when the smoothness of several functions needs to be estimated simultaneously.
Furthermore, the iterative nature is beneficial because the smoothness parameters can be initialized fairly large, leading to a very stable initial penalized fit. During the optimization they are typically reduced, leading to more flexibility, but for each new penalized fit, $\bm{\theta}$ can be initialized by its penalized estimate from the previous iteration. This strategy makes the procedure both efficient and relatively robust against convergence to local optima of the penalized likelihood.
For a more detailed explanation see \citet{koslik2024efficient}.

For practical implementation, we only specify the penalized negative log-likelihood function in \texttt{R} using the forward algorithm and \texttt{penalty} function implemented in the \texttt{R} package \texttt{LaMa} \parencite{koslikLaMa2025}. 
We then use the \texttt{qreml()} function from the same package to jointly estimate the model and smoothness parameters. Internally, this function uses the automatic differentiation machinery provided by the \texttt{R} package \texttt{RTMB} \parencite{kristensen2025rtmb} to enable rapid evaluation of \eqref{eqn:penalized_llk} and its gradient based on a C++ tape.
\vspace{-\baselineskip}

%% file: content/03_simulation.tex
\section{Simulation Study}
\label{sec:simulation}
\vspace{-\baselineskip}
To assess the performance of the proposed methodology, we conducted the following simulation study. The latent state sequence $\{S_t\}_{t=1,\ldots,T}$ is generated from a 2-state time-non-homogeneous Markov chain with initial state distribution $\boldsymbol{\delta}^{(1)}$ = (0.5, 0.5). The transition probabilities depend on a uniformly distributed covariate \( z_{1,t} \sim \mathcal{U}(-1,1) \), such that the t.p.m.\ at time \( t \) is given by
\[
\boldsymbol{\Gamma}_{t} = 
\begin{pmatrix}
\gamma^{(t)}_{1,1} & \gamma^{(t)}_{1,2} \\
\gamma^{(t)}_{2,1} & \gamma^{(t)}_{2,2}
\end{pmatrix},
\]
each row being modeled using a multinomial logit specification with linear and quadratic covariate effects:
\[
\text{logit}\bigl( \gamma_{i,j}^{(t)} \bigr) = \eta_{i,j}(z_{1,t}),
\quad \text{where} \quad
\begin{aligned}
\eta_{1,2}(z_{1,t}) &= -1.8 + 1.5\,z_{1,t} - 2\,z_{1,t}^2, \\
\eta_{2,1}(z_{1,t}) &= -2.1 - 2\,z_{1,t} - 1\,z_{1,t}^2,
\end{aligned}
\]
and $\eta_{i,i}^{(t)}(z_{1,t}) = 0$ for identifiability.
The observations \( Y_t \) are conditionally normally distributed given the latent state \( S_t \), with state-specific location and scale parameters modeled as functions of a second covariate \( x_{1,t} \sim \mathcal{U}(-1,1) \), with
\( Y_t \mid \{ S_t = i \} \sim \mathcal{N}\!( \mu_t^{(i)}, (\sigma_t^{(i)})^2 ), \)
where the predictors for the conditional mean and the log-standard deviation are specified as
\begin{align*}
\mu_t^{(1)} &= x_{1,t} + x_{1,t}^2, 
& \log\bigl(\sigma_t^{(1)}\bigr) &= -0.5 + x_{1,t}^2, \\
\mu_t^{(2)} &= -1 + x_{1,t} + 0.5 \sin(\pi x_{1,t}), 
& \log\bigl(\sigma_t^{(2)}\bigr) &= \log(1 + 0.5 x_{1,t}).
\end{align*}

\noindent In each of $R=200$ independent simulation runs, we generated a time series of length \(T = 4{,}000\). 
All models were fitted via penalized maximum likelihood as described in Section~\ref{sec:2.2}.

\begin{figure}[t!]
    \centering
    \vspace{-0.8cm}
    \includegraphics[width=0.49\textwidth]{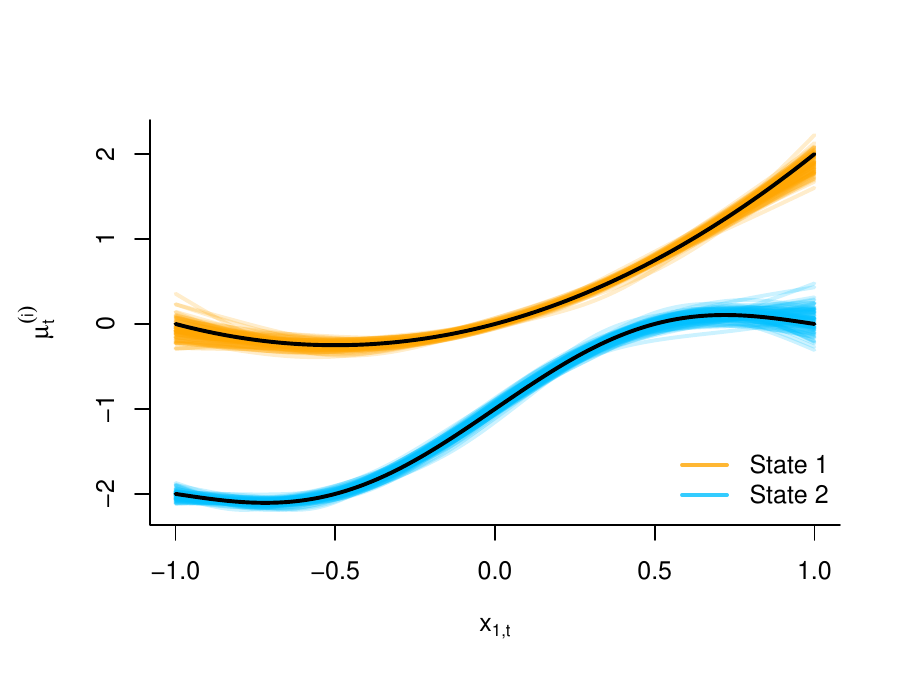}
    \includegraphics[width=0.49\textwidth]{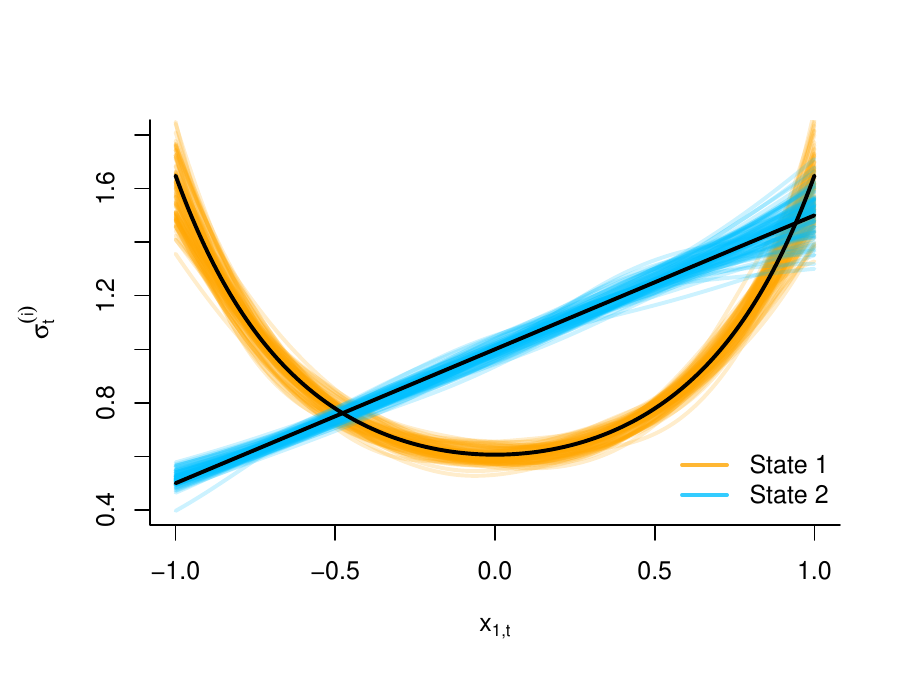}\par
\vspace{-0.8cm}
    \includegraphics[width=0.5\textwidth]{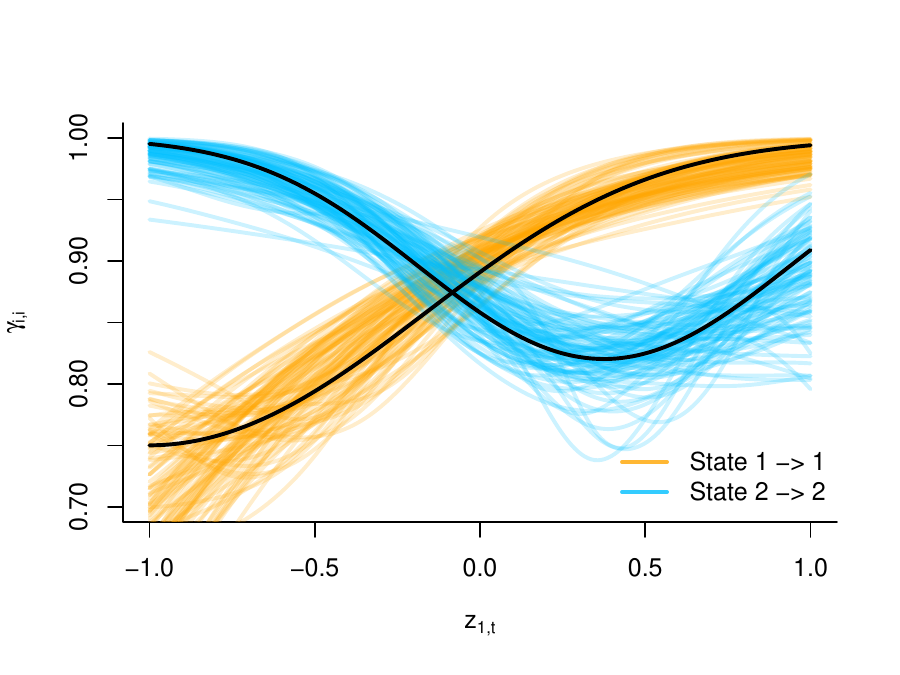}
    \caption{\textbf{Estimated state-dependent effects and transition probabilities.} 
    Displayed are the estimated state-dependent functions for the mean (top-left panel), the log-standard deviation (top-right panel), and the transition probabilities (bottom panel), obtained from 200 simulation runs. The estimates for States 1 and 2 are shown in orange and blue, respectively. The true functions are indicated by solid black lines. The bottom panel visualizes the estimated probabilities of remaining in States 1 and 2     along with the corresponding true curves.}
    \label{fig:sim_results}
\end{figure}

The simulation results are visualized in Figure~\ref{fig:sim_results}. The estimated smooth functions for the conditional mean \(\hat\mu_t^{(i)}\), the log-standard deviation \(\log(\hat\sigma_t^{(i)})\), and the transition probabilities \(\hat\gamma^{(t)}_{i,i}\) closely recover the true functional relationships of the data-generating process. 
For the location and scale parameters, the estimates exhibit low variability and closely follow the true functions across the entire covariate domain.
In particular, the state-dependent mean \(\hat\mu_t^{(i)}\) and log-standard deviation \(\log(\hat\sigma_t^{(i)})\) functions are well identified in both regimes, with slightly increased variability near the covariate boundaries but overall close alignment with the truth.
Regarding the estimated transition probabilities, the functions \(\hat\gamma^{(t)}_{1,1}\) and \(\hat\gamma^{(t)}_{2,2}\) approximate the true transition patterns well, with only minor deviations near the boundaries of the covariate domain where data are sparse. The estimated smooths capture the covariate effects on the transition mechanism, demonstrating that the proposed approach can recover covariate-dependent switching behavior as well as the underlying distributional structure and latent dynamics.

%% file: content/04_application.tex
\section{Empirical Application}
\label{sec:application}
\vspace{-\baselineskip}
This section illustrates the application of the proposed non-homogeneous MS-GAMLSS framework through two case studies. The first example analyzes the dynamics of daily closing prices of the Lufthansa stock. The second one uses Spanish energy price data to demonstrate the flexibility of the modeling approach across distinct economic settings.
\vspace{-\baselineskip}
\subsection{Lufthansa Stock Prices} \label{sec:Lufthansa}
The analysis is based on 6{,}515 daily observations of Lufthansa’s stock price from August 1, 2007, to June 27, 2025. Lufthansa is one of the largest airline companies in Europe and is a constituent of the German DAX index. The sample period spans several major financial crises, including the global financial crisis (2008), the European sovereign debt crisis (2010), the COVID-19 pandemic downturn (2020), and subsequent episodes of pronounced market volatility. These events offer a rich setting for examining regime dynamics across distinct market environments.

The choice of covariates follows well-documented relationships between airline stock performance, commodity prices, and macroeconomic conditions. The Brent crude oil price, represented by the Brent crude oil futures contract, is included as the sole covariate affecting both the conditional mean and variance of returns, since fuel costs constitute a major component of airline operating expenses. Variations in oil prices can therefore have a direct and substantial impact on airline profitability \parencite{LufthansaHedging2024}.
In addition, the term spread (the difference between long- and short-term interest rates) serves as a covariate in the hidden state process to account for regime-switching dynamics driven by macroeconomic conditions. A narrow or inverted spread often signals increased recession risk, whereas a wider spread generally indicates expectations of economic growth \parencite{Estrella1991}. In the present application, the term spread is included with a lag of 360 days to account for the typical delay with which macroeconomic conditions, such as recessions, become observable following changes in the yield curve.

The model specification for Lufthansa closing prices is as follows:
\begin{align}
\text{LH Closing Price}_t \mid \{S_t = i\} &\sim \text{N}\!\left(\mu_t^{(i)}, \, \sigma_t^{(i)}\right), \notag\\
\mu_t^{(i)} &= \beta_{0,\mu}^{(i)} + s_{1,\mu}^{(i)}\!\big(\text{oil\_price}_t\big), \notag\\
\log(\sigma_t^{(i)}) &= \beta_{0,\sigma}^{(i)} + s_{1,\sigma}^{(i)}\!\big(\text{oil\_price}_t\big), \notag\\[1.2ex]
\text{logit}\!\big(\gamma_t^{(i,j)}\big) &= \beta_{0,\gamma}^{(i,j)} + s_{1,\gamma}^{(i,j)}\!\big(\text{spread}_{t-360}\big), 
\label{eq:model_spec_lh_skew}
\end{align}
with $i, j = 1,2$ and $i \neq j$. Here, $LH\;Closing\;Price_t$ denotes the daily closing price of Lufthansa stock, $\textit{oil\_price}_t$ the crude oil price, and $\textit{spread}_{t-360}$ the term spread from one year ago. The smooth functions $s_{\cdot}^{(i)}(\cdot)$ allow for nonlinear covariate effects on the state-dependent mean and scale, while the transition probabilities are modeled via a (state-specific) logit link as functions of the term spread.
\newline

\begin{figure}[t]
\vspace{-1.3cm}
    \centering  \includegraphics[width=0.49\linewidth]{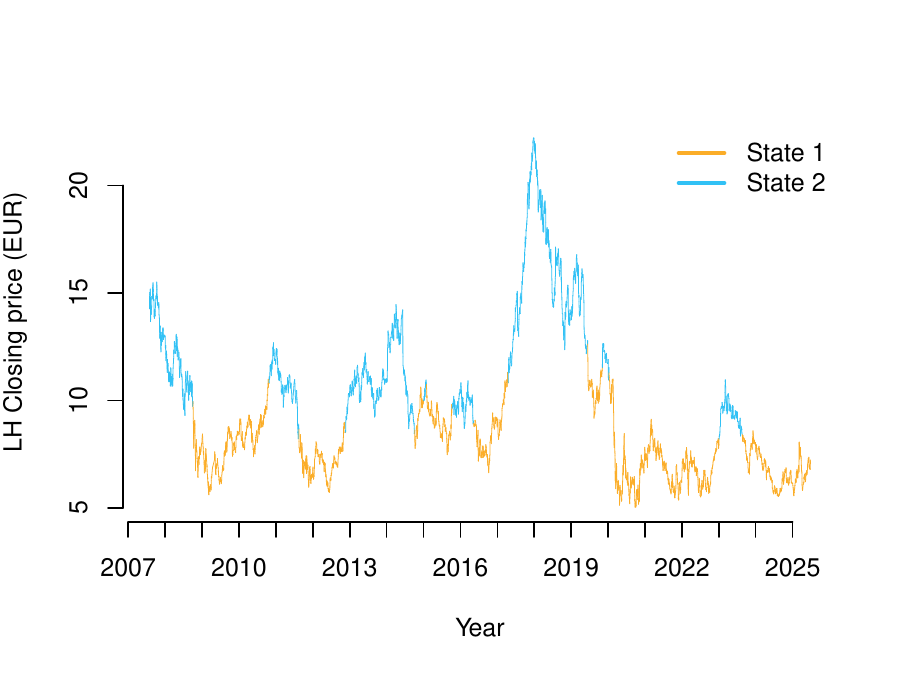}
    \includegraphics[width=0.49\linewidth]{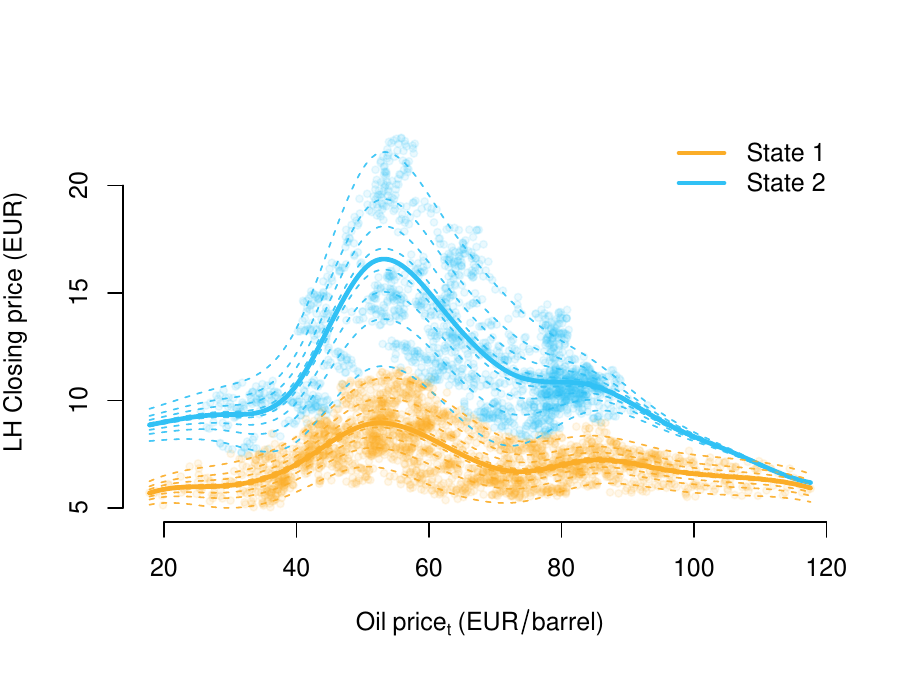}
    \vspace{-0.3cm}
    \caption{\textbf{Decoded time series of closing prices and estimated covariate effects.} 
    Left panel: Decoded time series under the fitted non-homogeneous MS-GAMLSS. Observations are colored by the most probable latent state.
    Right panel: Estimated state-dependent conditional means of closing prices as a function of the crude oil price.  
    Solid lines represent the estimated conditional expectation in each state. 
     Dashed lines show conditional quantiles of the fitted normal distribution across a regular grid of probability levels (5\%–95\%).
     }
    \label{fig:decoded_lh}
    \vspace{-0.3cm}
\end{figure}

\vspace{-\baselineskip}
The decoded time series in Figure~\ref{fig:decoded_lh} reveals two distinct regimes. State~1 (orange) corresponds to periods of lower closing prices and more stable pricing, while State~2 (blue) reflects higher and greater variation. The state sequence was obtained using the Viterbi algorithm \parencite{Viterbi1967}, which finds the most likely sequence of hidden states given the observed data and the estimated model parameters.

\begin{figure}[t]
\vspace{-0.6cm}
    \centering
    \includegraphics[width=0.49\linewidth]{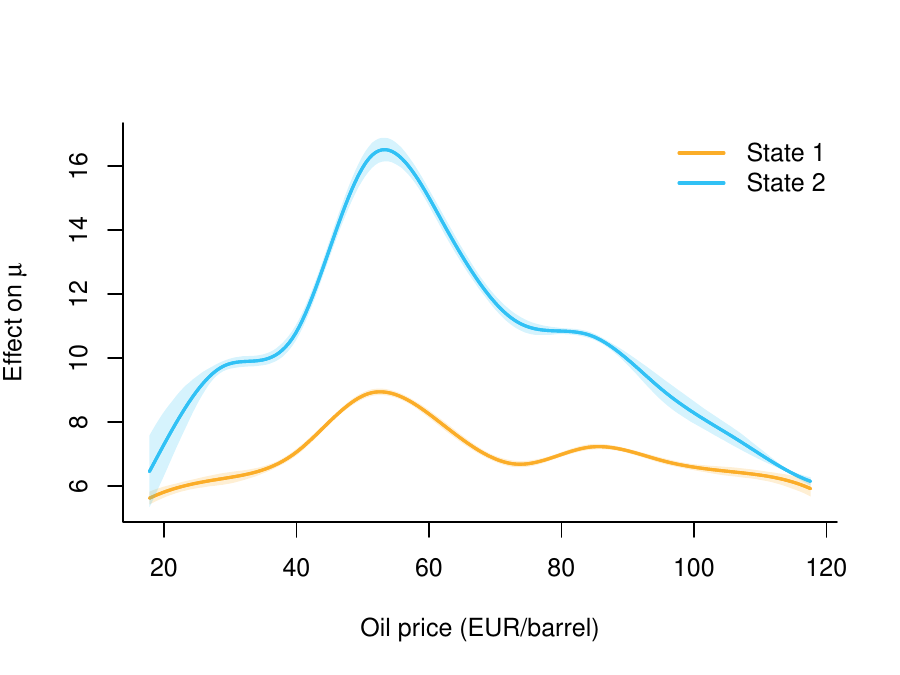}
    \includegraphics[width=0.49\linewidth]{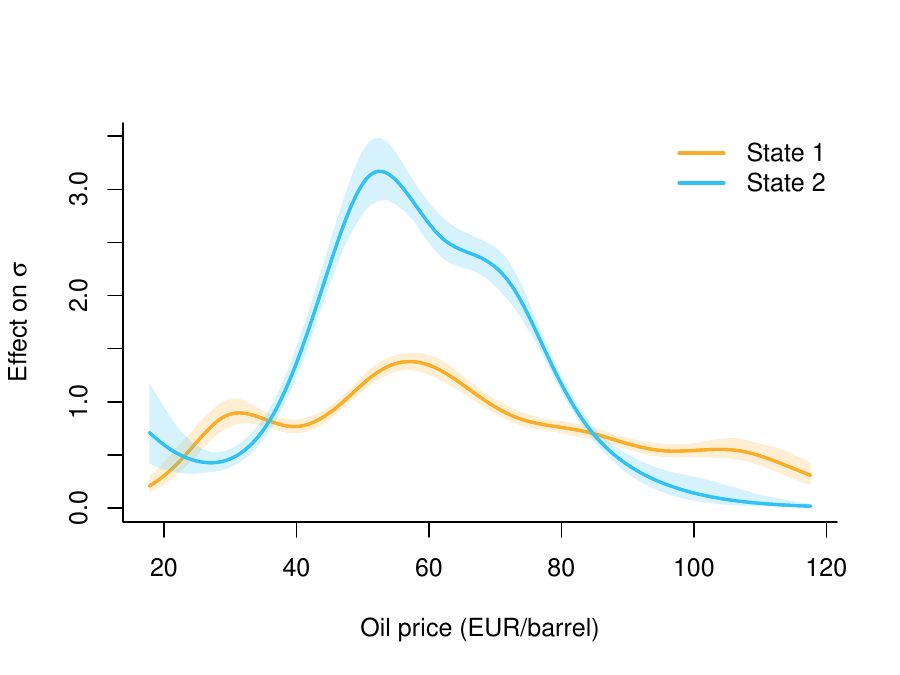}
    \caption{\textbf{Estimated state-dependent effects of the oil price.} 
    Effects on the conditional mean (left) and standard deviation (right) of the daily closing prices of Lufthansa stock. 
    Lines represent the estimated state-specific predictors, while shaded areas denote 95\% pointwise confidence intervals obtained from the posterior distribution of the state-dependent parameters in the fitted 2-state normal MS-GAMLSS.}
    \label{fig:lh_effectplots}
\end{figure}

The estimated smooth effects of the crude oil price (\textit{oil\_price}$_t$) on the conditional parameters are shown in Figure~\ref{fig:lh_effectplots}. In both states, the conditional mean of the daily closing prices initially increases sharply with rising oil prices. State~2 is generally associated with a higher conditional mean level across the oil price range, peaking around a closing price of 16~EUR when oil prices are close to 60~EUR per barrel. In contrast, State 1 remains on a lower level, fluctuating around 7~EUR on average.
For the conditional standard deviation, values in State~1 remain predominantly in the lower half of the $\sigma$ scale. In contrast, State~2 exhibits a generally stronger influence on volatility, particularly in the mid-range of oil prices, where the effect on $\sigma$ can approximately double.

\begin{figure}[t]
\vspace{-0.8cm}
    \centering
    \includegraphics[width=0.49\textwidth]{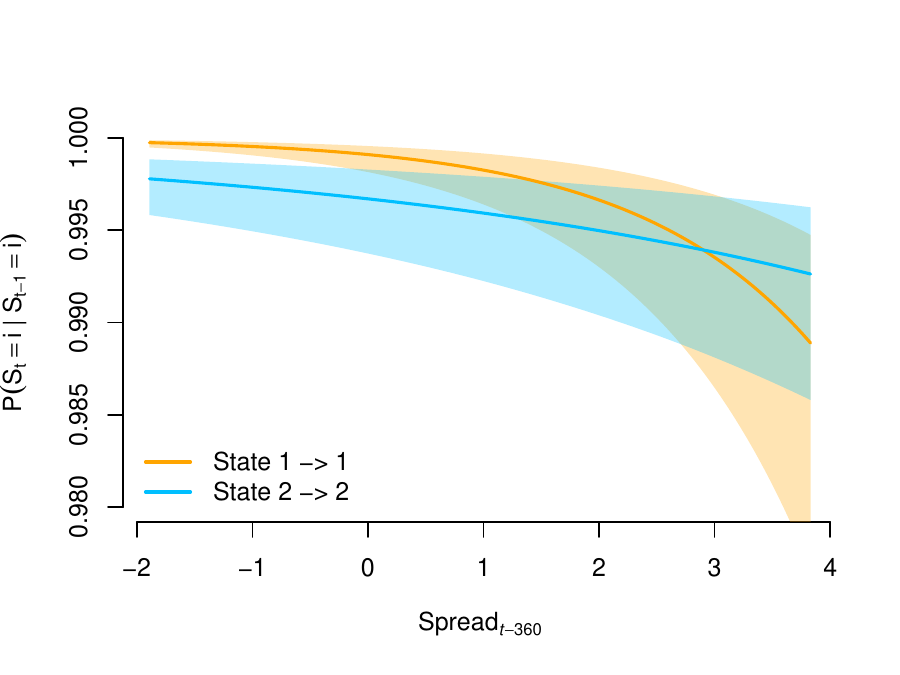}
    \includegraphics[width=0.49\textwidth]{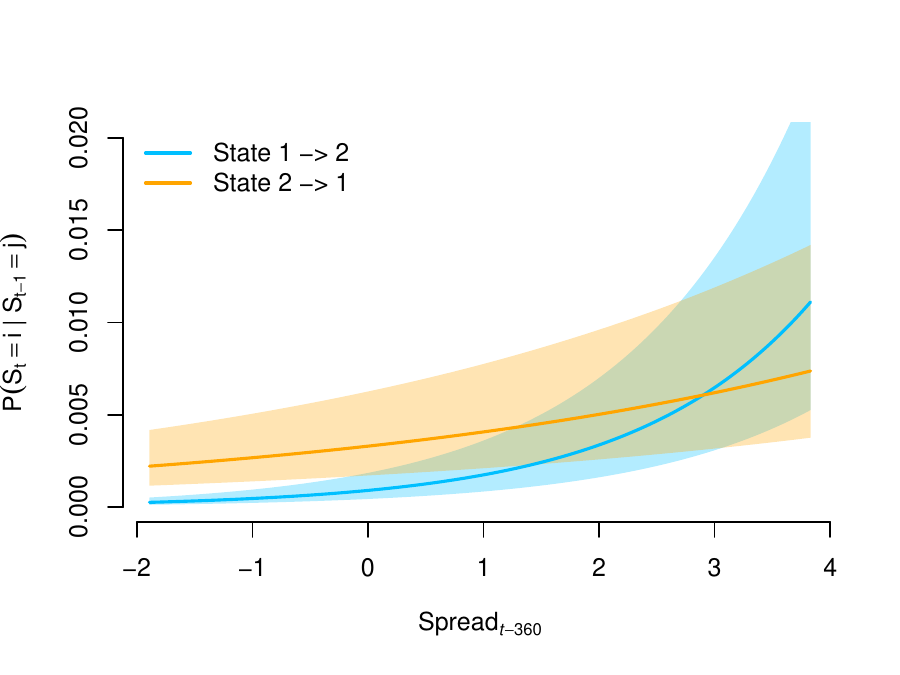} \par
\vspace{-0.8cm}
    \includegraphics[width=0.5\textwidth]{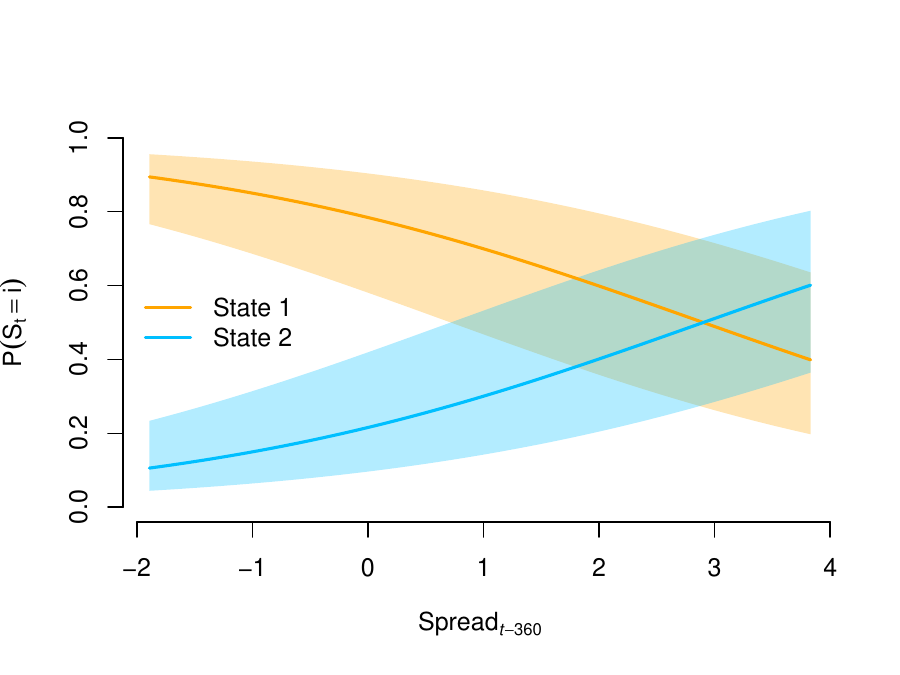}
    \caption{\textbf{Estimated transition probabilities as functions of the lagged term spread.} Left: Probabilities of remaining in the same regime. Right: Probabilities of switching between regimes. Bottom: Covariate-dependent stationary state probabilities as functions of the lagged term spread. Shaded areas denote 95\% pointwise confidence intervals derived from the posterior distribution of the covariate-dependent transition probabilities.}
    \label{fig:trans_probs}
    \vspace{-0.4cm}
\end{figure}

Figure~\ref{fig:trans_probs} illustrates the estimated state transition probabilities as functions of the lagged term spread. The left panel shows the probabilities of remaining in the current regime, while the right panel displays the probabilities of switching between states. Shaded areas represent pointwise 95\% confidence intervals for the stationary distribution, computed from posterior samples of the transition model parameters based on the observed Fisher-Information matrix, where the stationary distribution is evaluated for each draw and summarized by empirical quantiles. Solid lines show the estimated transition probabilities based on the maximum penalized likelihood estimates. Higher spreads correspond to decreased persistence in State~1, whereas lower spreads are associated with less frequent transitions to State~2. 

The bottom of Figure~\ref{fig:trans_probs} shows how the stationary distribution varies with the spread, given the estimated transition probabilities. When the spread is negative, the probability of being in State~1 is higher, which aligns with previous results indicating that State~1 corresponds to a lower price level. This probability decreases as the spread increases, until at a spread slightly below 3 the probabilities intersect and State~2 becomes more likely. This pattern is consistent with the interpretation that higher spreads are associated with positive economic expectations. 

The relatively wide confidence intervals reflect the fact that the decoded state sequence (see Figure \ref{fig:decoded_lh}) exhibits only a small number of likely regime transitions, which limits the amount of information available for estimating spread-dependent transition behavior.

\begin{figure}[H]
\vspace{-0.3cm}
    \centering
    \includegraphics[width=\textwidth]{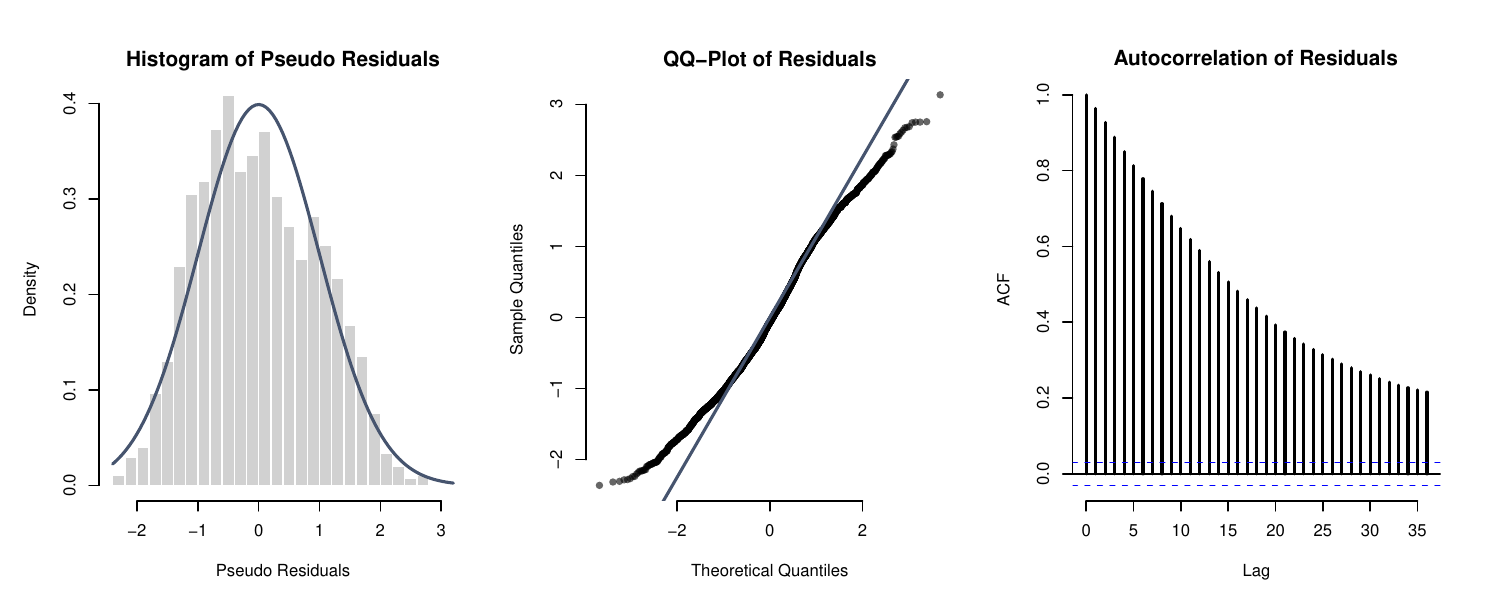}
    \caption{\textbf{Residual diagnostics for the fitted Lufthansa model.} 
    Left: Histogram of pseudo-residuals with the standard normal density overlaid, indicating an overall reasonable agreement with the theoretical distribution.
    Middle: QQ-plot of the pseudo-residuals, highlighting deviations in the extreme quantiles.
    Right: Autocorrelation function of the pseudo-residuals, showing remaining residual autocorrelation, which is expected given the absence of explicit autoregressive components in the model.
    }
    \label{fig:lh_pseudo}
    \vspace{-0.6cm}
\end{figure}

To assess the adequacy of the fitted model, we examine pseudo-residual diagnostics, shown in Figure~\ref{fig:lh_pseudo}. The pseudo-residuals are constructed via the probability integral transform and should therefore follow a standard normal distribution if the model is correctly specified. The histogram suggests an overall good fit in the tails, while indicating a slight shift of the central mass relative to the standard normal distribution. The QQ-plot, however, shows noticeable deviations in the extreme quantiles, suggesting that the normal state-dependent distribution does not fully accommodate the observed asymmetry.

The autocorrelation function of the pseudo-residuals indicates remaining serial dependence. This outcome is expected, as the model does not include autoregressive components or other explicit mechanisms to capture short-term temporal dependence. The observed residual autocorrelation therefore does not indicate a misspecification of the conditional distribution. Although incorporating autoregressive structures would be straightforward, this extension is omitted in order to keep the model specification parsimonious and to maintain the focus on distributional modeling rather than additional dynamic complexity.
This pattern motivates the use of a slightly more flexible state-dependent distribution in the following application, where skewness is modeled explicitly using a skew-normal specification.
\vspace{-\baselineskip}
\subsection{Energy Prices in Spain}\label{sec:energy_prices}

As a second application, we analyze the daily average electricity price in Spain to demonstrate our modeling framework in a different economic setting. Following previous studies on this dataset \parencite{Sanchez2009, langrock2017markov, Adam2022, koslik2024efficient}, our focus is on modeling conditional distributional dynamics and regime-switching behavior driven by relevant economic covariates. The data, available in the \texttt{R} package \texttt{MSwM} \parencite{Sanchez2014}, comprise 1,760 business days from February 4, 2002, to October 31, 2008.

To retain a simple and interpretable model structure, the conditional mean, scale, and skewness within each state are modeled as functions of the oil price alone. In addition, the EUR-USD exchange rate enters the hidden state process, allowing regime transitions to be linked to exchange rate fluctuations. In this specification, oil prices capture global commodity market dynamics, while exchange rate movements reflect broader macroeconomic influences that drive shifts in the energy price regimes.

The 2-state MS-GAMLSS for the Spanish energy prices is specified as follows:
\begin{align}
\text{Energy~Price}_t \mid \{S_t = i\} &\sim \text{SN}\!\left(\mu_t^{(i)}, \, \sigma_t^{(i)}, \, \nu_t^{(i)}\right)\notag,\\
\mu_t^{(i)} &= \beta_{0,\mu}^{(i)} + s_{1,\mu}^{(i)}(\text{oil\_price}_t)\notag,\\
\log(\sigma_t^{(i)}) &= \beta_{0,\sigma}^{(i)} + s_{1,\sigma}^{(i)}(\text{oil\_price}_t),\\
\nu_t^{(i)} &= \beta_{0,\nu}^{(i)} + s_{1,\nu}^{(i)}(\text{oil\_price}_t)\notag,\\[1ex]
\text{logit}\bigl(\gamma_{i,j}^{(t)}\bigr) &= \beta_{0,\gamma}^{(i,j)} + s_{1,\gamma}^{(i,j)}(\text{exchange\_rate}_t)\notag,
\label{eq:model_spec_energy_skew}
\end{align}
with $i, j = 1,2$ and $i \neq j$. $ Energy~Price_t $ denotes the daily average electricity price in Spain, $\textit{oil\_price}_t$ the crude oil price, and $\textit{exchange\_rate}_t$ the exchange rate between Dollar and Euro. Here, we assume a skew-normal distribution for $ Energy~Price_t $ and fitted a 2-state MS-GAMLSS with state-dependent predictors for the conditional mean, conditional variance, and conditional skewness. The smooth functions $s^{(i)}(\cdot)$ capture non-linear covariate effects on the state-dependent parameters, while the transition probabilities are modeled via a logit link as state-specific functions of the exchange rate.

\begin{figure}[t]
\vspace{-1.4cm}
    \centering
        \includegraphics[width=0.49\linewidth]{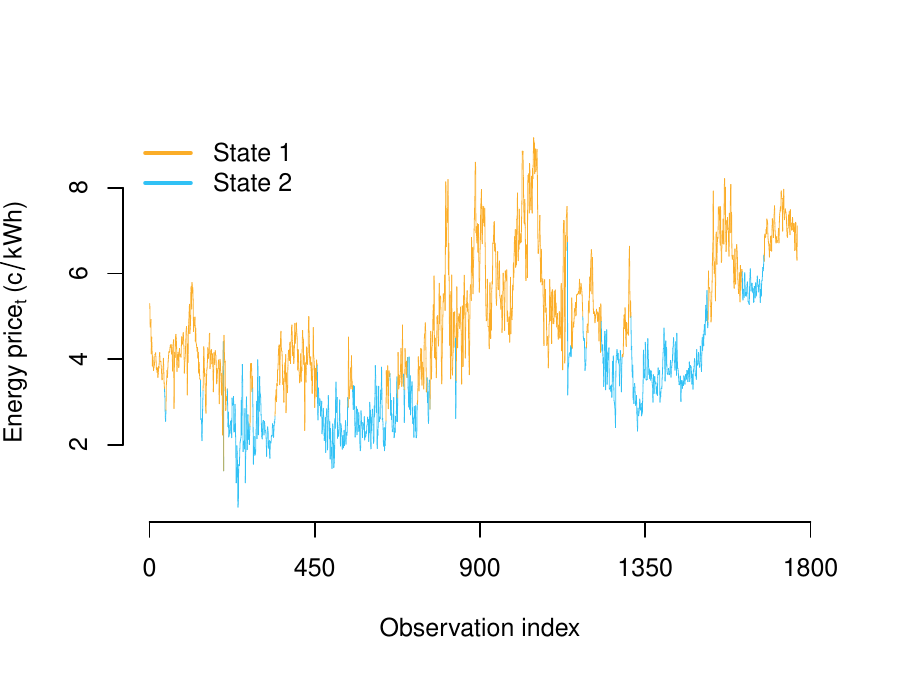}
    \includegraphics[width=0.49\linewidth]{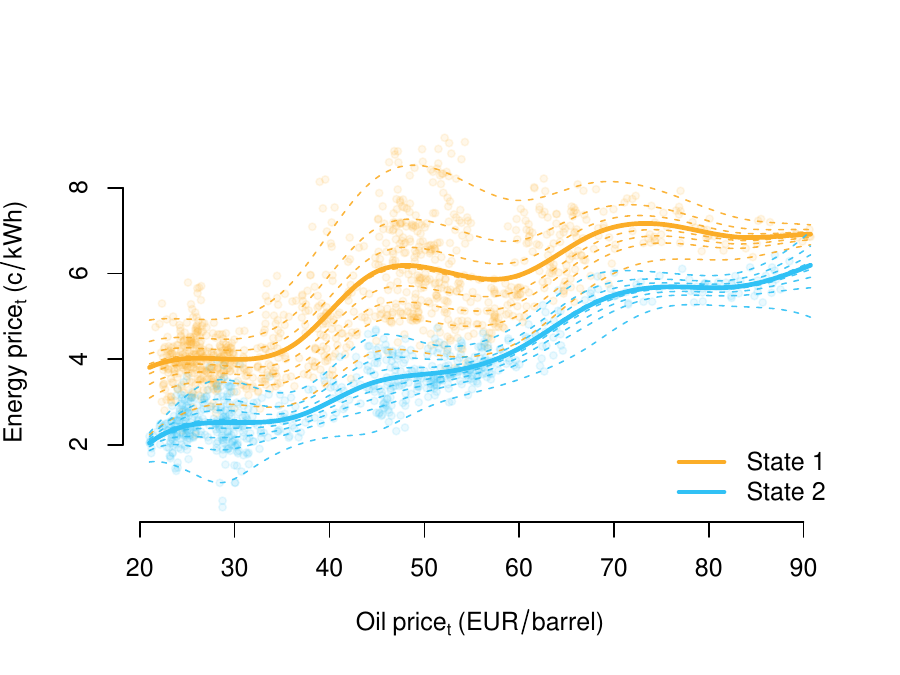}

    \caption{\textbf{Decoded time series of energy prices and estimated covariate effects.} 
    Left panel: Decoded time series under the fitted non-homogeneous MS-GAMLSS, with observations are colored by the most probable latent state.
    Right panel: Estimated state-dependent conditional means of energy prices as a function of the oil price. 
    Solid lines depict the estimated conditional expectation in each state, while  
     dashed lines show conditional quantiles of the fitted skew-normal distribution across a regular grid of probability levels (1\%–99\%).
}
    \label{fig:decoded_energy}
    \vspace{-0.4cm}
\end{figure}

The decoded time series in Figure~\ref{fig:decoded_energy} reveals two distinct regimes. State~1 (orange) corresponds to periods of higher electricity prices and greater variation, while State~2 (blue) reflects lower and more stable prices.

\begin{figure}[H]
\vspace{-1cm}
    \centering
    \includegraphics[width=0.49\linewidth]{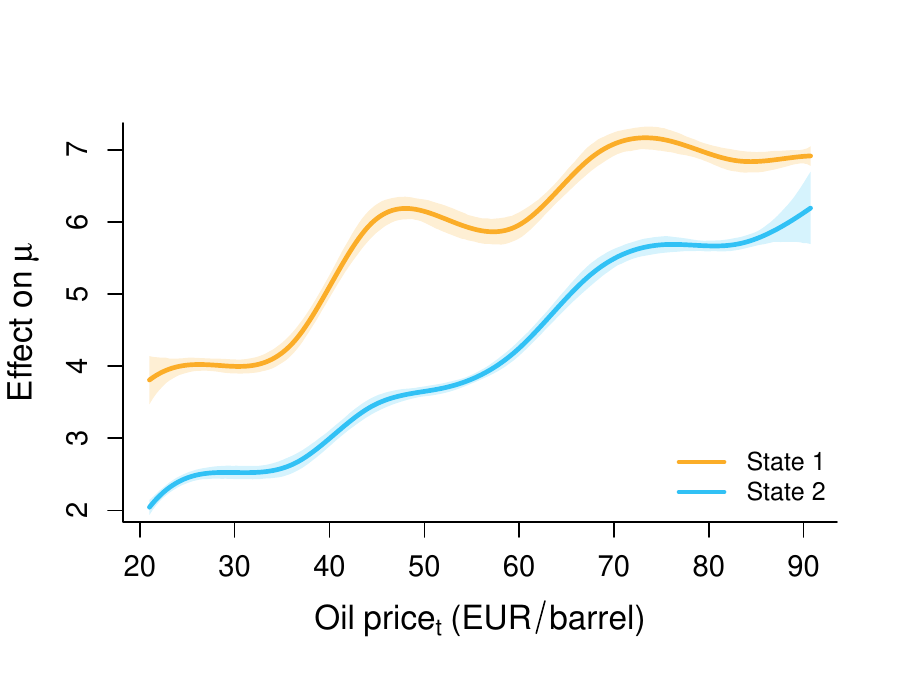}
    \includegraphics[width=0.49\linewidth]{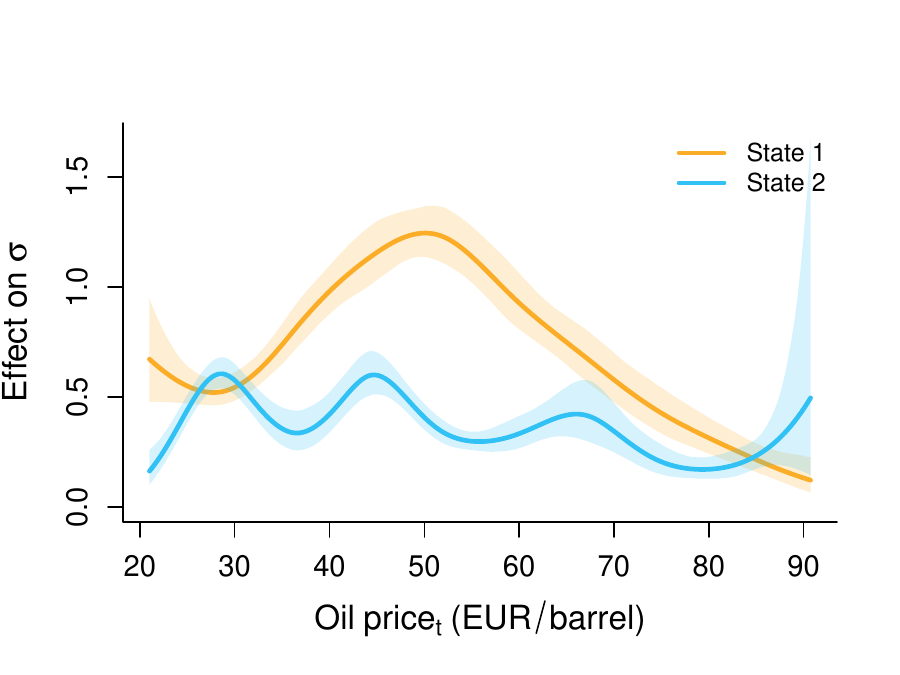} \par
\vspace{-0.8cm}
    \includegraphics[width=0.49\linewidth]{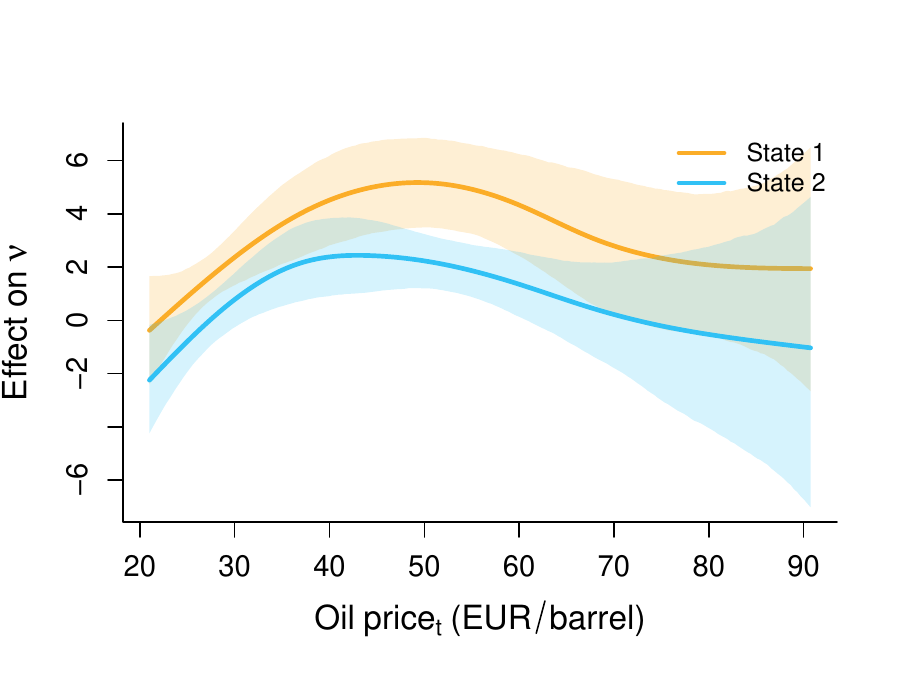}
\vspace{-0.3cm}
    \caption{\textbf{Estimated state-dependent smooth effects of the crude oil price.}
     Top-left: Effects on the conditional mean of daily energy prices. 
     Top-right: Effects on the conditional scale. 
     Bottom: Effects on the conditional skewness.   Lines show the estimated state-specific predictors; shaded areas indicate 95\% pointwise confidence intervals derived from the posterior distribution of the state-dependent parameters in the fitted two-state skew-normal MS-GAMLSS.
}
    \label{fig:energy_effectplots}
    \vspace{-1.3cm}
\end{figure}

Figure~\ref{fig:energy_effectplots} shows the estimated smooth effects of oil price on the conditional mean (top-left), standard deviation (top-right) and skewness (bottom) of the electricity price distribution, separately by state. 
    Both states exhibit a positive and non-linear association between oil price and electricity price. State~1 is associated with a higher conditional mean across the entire covariate domain and displays greater wiggliness in the estimated smooth. 
Clear differences across states also emerge for the conditional standard deviation. Volatility is elevated when \( 40 \leq \text{Oil~Price}_t < 60 \). This pattern is consistent with the decoded state sequence, where the highest conditional variance coincides with a high probability of being in State~1.
    The effect on the skewness parameter is state-dependent and non-linear. At low to medium oil prices, both states exhibit increasing skewness, with consistently larger values in State~1, indicating a more pronounced right tail of the electricity price distribution. As oil prices increase further, the skewness effect in State~1 levels off and gradually declines, while State~2 shifts towards negative skewness, suggesting a transition to heavier lower tails at high oil price levels.

\begin{figure}[t]
    \centering
    \includegraphics[width=0.49\textwidth]{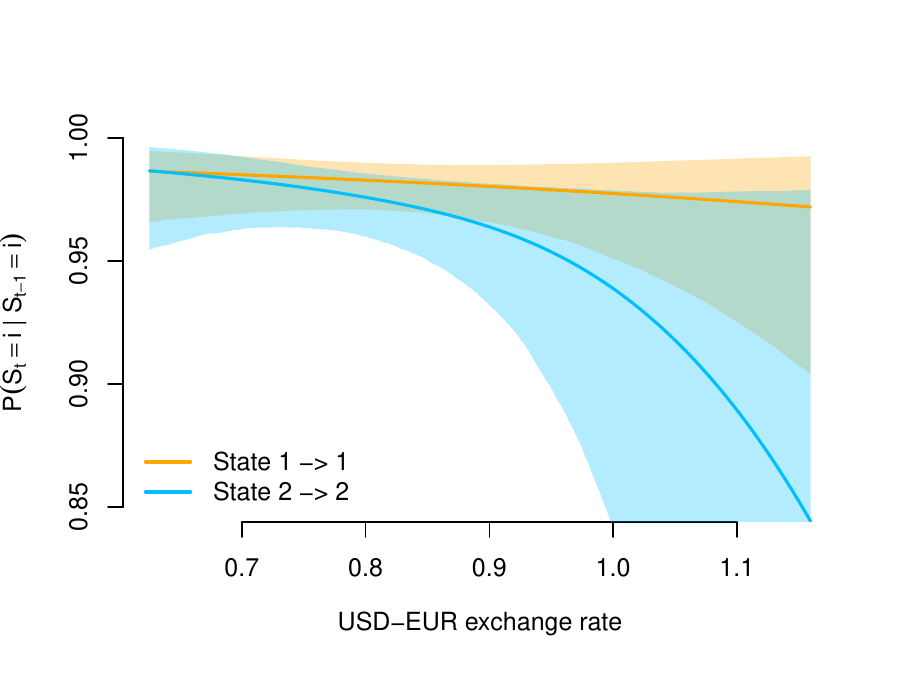}
    \includegraphics[width=0.49\textwidth]{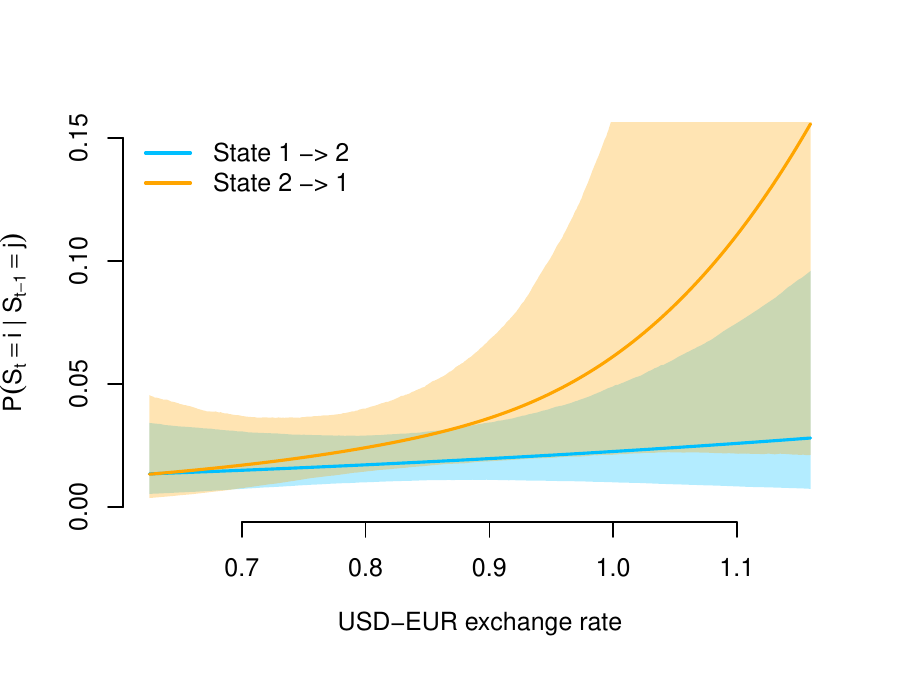}\par
\vspace{-0.8cm}
    \includegraphics[width=0.49\textwidth]{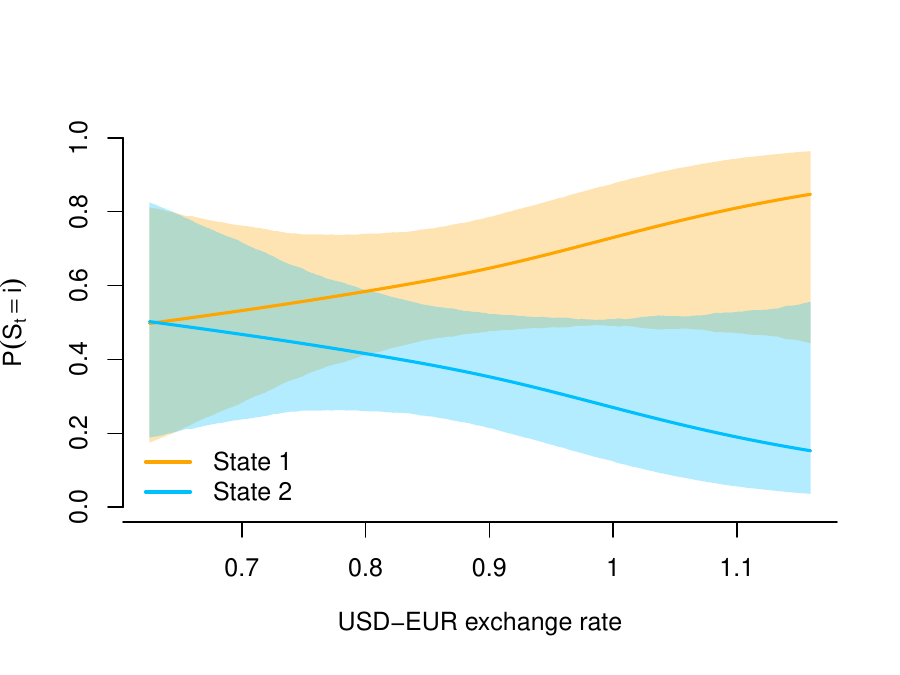}
    \caption{\textbf{Estimated transition probabilities as functions of the exchange rate.} Top-left: probabilities of remaining in the same regime. Top-right: probabilities of switching between regimes. Bottom: covariate-dependent stationary state probabilities as functions of the exchange rate. Shaded areas represent 95\% pointwise confidence intervals obtained from the posterior distribution of the covariate-dependent transition probabilities.}
    \label{fig:energy_trans_probs}
    \vspace{-0.3cm}
\end{figure}

Figure~\ref{fig:energy_trans_probs} presents the estimated transition probabilities as functions of the USD-EUR exchange rate. The left panel shows the probabilities of remaining in the current regime, while the right panel displays the probabilities of switching between states. The shaded areas represent 95\% confidence intervals derived from the estimated transition models. 
As the exchange rate increases, persistence in both states decreases. However, the probability of transitioning from State~2 to State~1 rises, leading to a higher likelihood of entering State~1 at higher exchange rate levels.

The covariate-dependent stationary distribution (bottom panel) shows a clear shift in regime allocation with increasing exchange rates. As the exchange rate rises, the stationary probability of State~1 steadily increases, while the probability of being in State~2 declines. Clearly, Figure~\ref{fig:energy_trans_probs} reveals that a higher exchange rate corresponds to an increased persistence in State~1, implying more stable periods of higher energy prices during times of currency depreciation. 

This pattern is consistent with the decoded state sequence (see Figure \ref{fig:decoded_energy}), which assigns State~1 to periods of relatively strong euro levels. The increasing persistence in State~1 suggests that elevated electricity prices tend to be more stable when the euro appreciates, potentially due to reduced import costs for energy commodities.

\begin{figure}[h!]
    \centering
    \includegraphics[width=\textwidth]{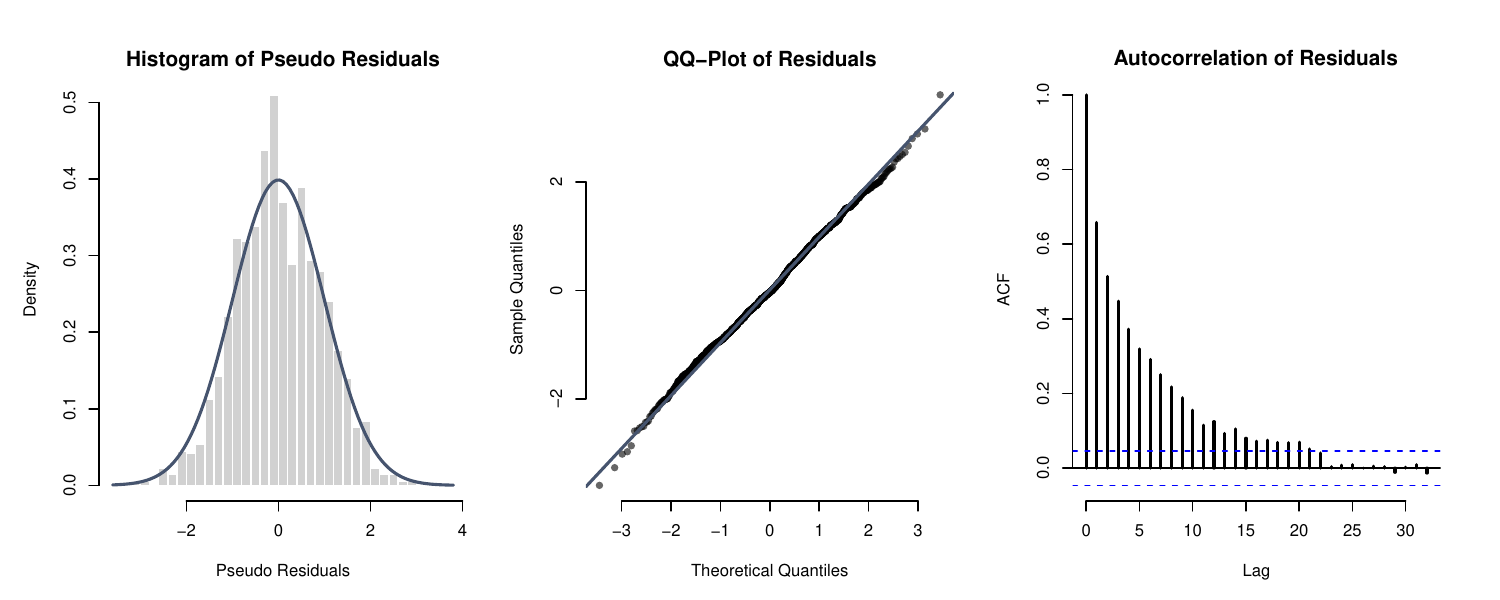}
    \caption{\textbf{Residual diagnostics for the fitted energy price model.} 
    Left: Histogram of pseudo-residuals with the standard normal density overlaid, indicating a close correspondence between the empirical and theoretical distributions.
    Middle: QQ-plot showing that the pseudo-residuals closely follow the standard normal quantiles, suggesting well-calibrated conditional response distributions.
    Right: Autocorrelation function of the pseudo-residuals, indicating remaining residual autocorrelation, which is expected since the model does not include explicit autoregressive components to capture temporal dependence.}
    \label{fig:energy_pseudo}
\end{figure}

To assess the adequacy of the fitted model, we examine the pseudo-residual diagnostics shown in Figure~\ref{fig:energy_pseudo}. The construction of these diagnostics follows the same principles as outlined in Section~\ref{sec:Lufthansa}.
 The histogram and QQ-plot indicate a close correspondence with the standard normal distribution, suggesting that the skew-normal state-dependent distribution provides a well-calibrated description of the marginal behavior.
The autocorrelation function shows remaining serial dependence in the pseudo-residuals, which is expected given the absence of explicit autoregressive components.

%% file: content/05_discussion.tex
\section{Discussion}
\label{sec:discussion}

We proposed a non-homogeneous MS-GAMLSS framework for modeling time series with regime-dependent distributional parameters, in which both the hidden state process and the state-dependent process may depend on covariates. Incorporating covariates into the transition process enables inference on the drivers of regime-switching dynamics. In particular, the estimated transition probabilities indicate how changes in explanatory variables influence the persistence of, or switching between, latent states. This, in turn, provides insight into the mechanisms underlying structural shifts in the data-generating process.
Our simulation experiment confirms that the proposed model can successfully recover regime-dependent covariate effects in both the state-dependent parameters and the transition probabilities. The empirical applications further demonstrate the economic relevance of these effects.

Although the empirical illustrations rely on normal and skew-normal response distributions, the GAMLSS framework accommodates a wide range of distribution families. These include distributions for continuous, count, and binary data, as well as skewed, heavy‑tailed, and zero‑inflated distributions that frequently occur in financial and economic applications. This flexibility makes the proposed framework applicable far beyond the examples considered here and suitable for diverse data structures and stylized features. Future research may exploit these distributional extensions to tailor the model more closely to the specific properties of empirical data.

In addition to general distributional modeling, the framework is well suited for estimating risk measures such as Value-at-Risk and Expected Shortfall, which are based on quantiles rather than expectations and therefore focus on the tails of the distribution—an aspect of particular importance in financial applications. Accurate tail estimation is challenging, as financial return series often exhibit heavy tails, skewness, and other departures from normality. MS-GAMLSS allows all distributional parameters to vary across regimes and with covariates, potentially improving tail estimation under changing market conditions. Although this was not the primary focus of the present study, future work could investigate the suitability of the model for short-term risk prediction in settings where full distributional forecasts are required (\citealp{rockafellar2001, acerbi2002}).

Further extensions could relax the assumption of a finite number of discrete regimes. In some applications, particularly in financial time series, a continuous latent state process --- as in state space models --- may be more appropriate. Allowing for continuous transitions between regimes would enable smoother adaptation to gradual changes in underlying market conditions and represents a promising direction for future research.
\vfill
\newpage

%% file: content/XX_Appendix.tex
\section{Standardized Forward Variables}
\label{sec:app_stand_forw_var}

Due to technical issues arising during the maximization process, such as numerical underflow or overflow, the likelihood can become extremely small or large for large time series $T$. As a result, the likelihood may be rounded to zero (underflow) or returned as $+\infty$ (overflow) in software such as \texttt{R}. This issue occurs because the likelihood is computed as a matrix product, where each forward variable $\boldsymbol{\alpha}_t$ is multiplied recursively, leading to exponentially growing or shrinking values. To address this, a scaling strategy is applied by introducing \textit{standardized forward variables}, which are normalized to avoid numerical instability. Let $\boldsymbol{\phi}_t = (\phi_t(1), \phi_t(2), \dots, \phi_t(N))$ represent the standardized forward variables at time $t$, defined as 
\begin{align*}
\boldsymbol{\phi}_t = \frac{\boldsymbol{\alpha}_t}{\boldsymbol{\alpha}_t \boldsymbol{1}^\top} = \frac{\boldsymbol{\alpha}_t}{\sum_{j=1}^N \alpha_t(j)}.
\end{align*}
The scaling ensures that the sum of the standardized forward variables equals one, i.e., $\sum_{j=1}^N \phi_t(j) = 1$, preventing the values from becoming arbitrarily large or small. The recursion for $\boldsymbol{\phi}_t$ is then updated as 
\begin{align*}
\boldsymbol{\phi}_t = \frac{\boldsymbol{\phi}_{t-1} \boldsymbol{\Gamma} \mathbf{P}(y_t)}{\boldsymbol{\phi}_{t-1} \boldsymbol{\Gamma} \mathbf{P}(y_t) \boldsymbol{1}^\top},
\end{align*}
where $\mathbf{P}(y_t)$ is the diagonal matrix of state-dependent response densities at time $t$, and $\boldsymbol{\Gamma}$ is the transition probability matrix.

To compute the log-likelihood $\ell(\boldsymbol{\theta})$ in terms of the standardized forward variables, we first observe that the likelihood can be expressed as the product of scaled values:
\begin{align*}
\mathcal{L}(\boldsymbol{\theta}) = \prod_{t=1}^T \frac{\boldsymbol{\alpha}_t \boldsymbol{1}^\top}{\boldsymbol{\alpha}_{t-1} \boldsymbol{1}^\top}.
\end{align*}
Taking the logarithm, the log-likelihood becomes
\begin{align*}
\ell(\boldsymbol{\theta}) = \log \big(\mathcal{L}(\boldsymbol{\theta})\big) = \sum_{t=1}^T \log \bigg(\frac{\boldsymbol{\alpha}_t \boldsymbol{1}^\top}{\boldsymbol{\alpha}_{t-1} \boldsymbol{1}^\top}\bigg),
\end{align*}
which simplifies to
\begin{align*}
\ell(\boldsymbol{\theta}) = \log \big(\boldsymbol{\delta}^{(1)} \mathbf{P}(y_1) \boldsymbol{1}^\top\big) + \sum_{t=2}^T \log \big(\boldsymbol{\phi}_{t-1} \boldsymbol{\Gamma} \mathbf{P}(y_t) \boldsymbol{1}^\top\big).
\end{align*}